\documentclass[sigconf,nonacm]{acmart}

\AtBeginDocument{%
  }

\usepackage{listings}
\usepackage{xcolor}
\usepackage{booktabs}
\usepackage[symbol]{footmisc}

\newcommand{\todo}[1]{{\textcolor{red}{\textbf{#1}}}}

\definecolor{codegray}{gray}{0.95}

\usepackage{subcaption} %
\usepackage{adjustbox}
\usepackage{xspace}
\usepackage{xcolor}
\usepackage{wrapfig}
\usepackage{tabularx}
\usepackage{booktabs}
\usepackage{ragged2e}  
\usepackage{multirow}
\usepackage{fancyvrb}
\usepackage{float}   %
\newfloat{listing}{htbp}{lst} %
\floatname{listing}{Listing}  %

\usepackage{listings}
\usepackage{xcolor}
\usepackage{textcomp} %

\definecolor{codepurple}{rgb}{0.58,0,0.82}
\definecolor{codered}{rgb}{0.8,0,0}
\definecolor{codegreen}{rgb}{0,0.6,0}
\definecolor{codewhite}{rgb}{1,1,1}

\newcommand{\code}[1]{\textcolor{codered}{\textasciigrave#1\textasciigrave}}

\newcommand{\authortodo}[2]{\todo{#1: #2}}

\newcommand{\sam}[1]{\authortodo{Sam}{#1}}
\newcommand{\Space}[1]{}
\newcommand{\Comment}[1]{}

\begin{document}

\title{Towards a Human-in-the-Loop Framework for \\ Reliable Patch Evaluation Using an LLM-as-a-Judge
}

\author{
 Sherry Shi, Renyao Wei, Michele Tufano, Jos{\'e} Cambronero, Runxiang Cheng, \\ Franjo Ivan\v{c}i\'{c}, Pat Rondon
}
\affiliation{%
  \institution{Google, USA}
  \country{}
}

\email{
{
sherryyshi,
renyaow,
tufanomichele, 
jcambronero, 
chengsam,
ivancic,
rondon
}@google.com}

\renewcommand{\shortauthors}{Google et al.}
\renewcommand{\shorttitle}{Towards a Human-in-the-Loop Framework for Reliable Patch Evaluation Using an LLM-as-a-Judge}

\begin{abstract}
Reliable evaluation is crucial for advancing Automated Program Repair (APR), but prevailing benchmarks rely on execution-based evaluation methods (unit test pass@k), which fail
to capture true patch validity.
Determining validity can require costly manual annotation.
To reduce this cost, we introduce a human-in-the-loop approach to LLM-based patch validity judgment.
Inspired by the observation that human judgment is better aligned when using a shared rubric, we first employ an LLM to generate a per-bug rubric, followed by a one-time human review and optional refinement to this rubric, and then employ an LLM to judge patches using the refined rubric.
We apply this approach to assign binary validity labels to patches for issues found by Google sanitizer tools.
Our results show that
this approach yields substantial agreement with human consensus (Cohen's kappa 0.75), high recall (0.94) and high precision (0.80), when considering patches that have unanimous agreement from 3 human raters on the validity labels.
On the full dataset including patches where human raters disagree, we find this approach can still be further improved (Cohen's kappa 0.57, recall 0.93, precision 0.65) and identify possible future directions.
\end{abstract}

\maketitle
\section{Introduction}
\label{sec:introduction}

Automated Program Repair (APR) stands as a promising frontier in software engineering, with the potential to significantly enhance developer productivity by automatically fixing bugs.
As LLM-based APR systems become more capable, their deployment in industrial settings has begun in earnest \cite{rondon2025evaluating, takerngsaksiri2025human, maddila2025agentic}.
The success of these systems hinges on a reliable evaluation procedure to ensure that offline results accurately reflect online performance, so that development and deployment decisions can be made with high confidence.

For years, APR research has employed execution-based evaluation as the de facto standard, with popular benchmarks (such as MBPP~\cite{austin2021program}, HumanEval~\cite{chen2021evaluating}, and SWE-Bench~\cite{jimenez2023swe}) measuring functional correctness through test pass rates like pass@k.
While instrumental, this evaluation paradigm has a critical limitation: a patch can contain additional behaviors that are not assessed by the tests, potentially invalidating it.
Indeed, \citet{yu2025utboost,wang2025solved,zhu2025establishing} uncovered a significant number of cases where patches that passed tests in benchmarks like SWE-Bench were nonetheless functionally incorrect.
These invalid patches highlight a gap between test performance and true validity.
While these patches can be caught by human review, manual patch assessment is time-consuming and expensive to scale.
More importantly, we find that manual patch assessment suffers from its own critical flaw: inconsistency. In our study of the human assessment process, we found low inter-rater agreement (Fleiss's kappa 0.31) among developers assessing patch validity independently.

However, such inconsistency can be largely resolved by allowing developers to perform their assessments with a shared, high-quality rubric as reference.
As we will show in \S\ref{sec:motivation}, inter-rater agreement  substaintially increases if all raters assess generated patches of a bug following the same set of fix requirements for that bug.
More importantly, such rubric-based evaluation paradigm allows us to consider the potential of employing LLM-as-a-Judge for offline APR patch assessment.

LLM-as-a-Judge has emerged as an increasingly attractive approach to overcome the scalability limitation of manual evaluation, where an LLM provides an assessment of the code generated by another model.
Although misjudgements by the LLM call its practicality into question, \citet{tong2024codejudge,crupi2025effectiveness,zheng2023judging} have shown promise in this direction. In particular, \citet{zhuge2024agent} achieved 90\% alignment rate between humans and their agentic LLM judge on AI development tasks. While their work demonstrates a high alignment rate with humans using a set of manually-crafted set of requirements, the practical challenge of crafting these requirements remains. The significant manual effort required to create high-quality evaluation criteria for diverse and evolving sets of bugs remains a critical bottleneck for scalable evaluation in APR, highlighting an opportunity for LLMs to assist.

To overcome the reliability and scalability limitations of human evaluation, especially for bugs with no pre-existing fix requirements, we propose a human-in-the-loop framework to evaluate APR patch validity using LLM-as-a-Judge. Building on our insight that a shared, high-quality patch validation rubric is the cornerstone of reliable evaluation, our two-stage framework operationalizes rubric-guided evaluation at scale. First, an LLM generates a draft rubric for a given bug, which two human experts then review and refine into a ``golden'' evaluation standard. Second, an LLM judge uses this golden rubric to assess the validity of candidate patches.
Our evaluation, conducted on 115 patches for 48 bugs reported by Google's sanitizer tools, demonstrates that this approach yields substantial agreement with human consensus (Cohen's kappa 0.75), high recall (0.94) and high precision (0.80), when considering patches that have unanimous human agreement from 3 raters on the validity labels (70.4\% of the dataset). On the full dataset including patches where human raters disagree, we find
this approach can still be further improved (Cohen's kappa 0.57, recall 0.93, precision 0.65) and we identify possible directions.

This paper makes the following contributions:
\begin{itemize}
    \item An empirical study of human evaluation for code patches that demonstrates low inter-rater agreement when raters act independently, but and improves when raters refer to a common rubric.
    \item A novel human-in-the-loop framework that leverages an LLM to generate task-specific rubrics which are then refined by a human developer.
    \item A large-scale evaluation demonstrating that LLM-as-a-Judge, when guided by human-refined rubrics, achieves moderate to substantial levels of agreement with humans.
\end{itemize}

\section{Motivation}
\label{sec:motivation}

The ultimate goal of APR is to generate correct patches. However, the field has long grappled with a significant ``evaluation gap'' between Fail-to-Pass (F2P) patches, those that make a failing test suite pass (measured by pass@k), and correct patches, those that actually address the underlying bug and would pass human code review (measured by valid@k). This gap is not a new phenomenon; it has persisted across generations of APR techniques~\cite{qi2015analysis}. For instance, on Defects4J v1.2~\cite{JustJE2014}, Recoder~\cite{zhu2022syntax} found that only 54.3\% of its plausible patches were correct, ContrastRepair~\cite{kong2024contrast} showed 62.5\% correctness (75 out of 120 F2P), and ChatRepair~\cite{Xia_2024} showed 51.5\% correctness (52 out of 101 F2P).

This discrepancy highlights that pass@k is an insufficient signal and it is limited by the quality of the underlying tests. Test suites may have insufficient coverage, weak assertions, or missing edge-case handling, making them susceptible to ``hacks'' where a patch overfits to the tests without addressing the bug's root cause. While one can potentially improve the execution-based signal by enhancing test quality, it is not a complete solution; in large-scale, complex software systems, it is often intractable to create a test suite that perfectly captures all functional and non-functional requirements.

Consequently, to bridge this evaluation gap and determine the final verdict on patch validity, the field has already turned to manual assessment from expert developers as the de facto gold standard. However, the reliability of this manual process is not universally established and may vary by context. For instance, \citet{oliva2025spice} found low inter-rater agreement when applying discrete labels for issue clarity and test coverage on complex, SWE-Bench-like software tasks.

As the first step towards building a more automated system to evaluate patch validity, we conduct a preliminary study to answer the following question: \textit{To what extent do human raters agree on the validity of APR patches, and how does the evaluation rubric affect their agreement?}

Our study involves three of the authors acting as raters. We used a preliminary dataset of 52 F2P patches generated for five distinct bugs reported by Google's Sanitizer tools.
For each bug, the three raters referred to the bug report and ground-truth patch to independently author their own evaluation rubric, which outlines the requirements for a valid fix. Each rater then produced a binary assessment (VALID or INVALID) for all 52 patches under two settings: (1) using self-authored rubric, and (2) using a rubric authored by another rater.

First, to quantify the reliability of manual patch validity assessment, we calculated Fleiss's kappa on the raters’ assessments based on their self-authored rubrics. Our analysis revealed a low level of agreement among the three raters (e.g., 0.31 as shown in Table~\ref{tab:self-authored}). The main reasons for disagreements were different interpretations of the root cause, disagreements on the acceptability of unnecessary changes, disagreements on non-functional requirements, and misunderstandings of the code. This suggests that human raters’ patch validity assessment in the current manual evaluation process can be unreliable, in part because human raters can easily develop and follow their own subjective criteria.

\begin{table}[t!]
  \centering
  \caption{On 52 F2P patches of 5 bugs, three raters show substantial disagreement when using self-authored rubrics (Cohen’s kappa ranges from 0.19 to 0.39).}
  \label{tab:self-authored}
  \begin{tabular}{@{}llc@{}}
    \toprule
    \textbf{Raters} & \textbf{Rubric} & \textbf{Fleiss's $\kappa$ or Cohen’s $\kappa$} \\
    \midrule
    Rater 1, 2, 3 & Self-authored rubric & 0.31 \\
    Rater 1 and 2 & Self-authored rubric & 0.37 \\
    Rater 2 and 3 & Self-authored rubric & 0.19 \\
    Rater 3 and 1 & Self-authored rubric & 0.39 \\
    \bottomrule
  \end{tabular}
\end{table}

Further analysis reveals that the rubric is a primary contributor to discrepancies among human raters. Rater self-agreement, which compares assessments from the same rater using two distinct rubrics (their own and another's), was found to be moderate. Its Cohen's kappa values are between 0.35 and 0.49 (Table~\ref{tab:self-agreement}). Conversely, when pairs of raters utilized the same rubric for evaluation, their agreement significantly improved, yielding Cohen's kappa values ranging from 0.53 to 0.84 (Table~\ref{tab:same-rubric}). This preliminary study supports the hypothesis that basing patch validity assessment on a common rubric facilitates reliable and reproducible assessments of APR patch validity.

\begin{table}[t!]
  \centering
  \caption{Rater’s self-agreement is moderate (0.35 to 0.49) when using two different rubrics.}
  \label{tab:self-agreement}
  \begin{tabular}{@{}llc@{}}
    \toprule
    \textbf{Rater} & \textbf{Rubric} & \textbf{Cohen’s kappa} \\
    \midrule
    Rater 1 & Rater 1 rubric \& Rater 3 rubric & 0.35 \\
    Rater 2 & Rater 2 rubric \& Rater 1 rubric & 0.36 \\
    Rater 3 & Rater 3 rubric \& Rater 2 rubric & 0.49 \\
    \bottomrule
  \end{tabular}
\end{table}

This finding raises a natural next question: \textit{What constitutes a good rubric?} Intuitively, a rubric should be precise and informative if it is to be shared among human raters to assess patch validity. To investigate what constitutes such a ``golden rubric,'' the three initial human raters collectively analyzed their disagreements and distilled a set of best practices into a standardized template.
Listing~\ref{lst:rubric-summary} shows an example rubric following this template.
This template specifies key sections including the root cause of the bug, a checklist of requirements for a valid fix, and concrete examples of acceptable and unacceptable solutions.

\begin{listing}[t!]
\centering
\caption{Illustrative rubric example.}
\label{lst:rubric-summary}

\begin{lstlisting}
@\textcolor{codepurple}{\textbf{\#\# Root Cause}}@
The bug is a @\code{use\_of\_uninitialized\_value}@ error. A new member, @\code{redacted\_property\_name}@, was added to @\code{RedactedClass}@ in @\code{some/redacted/path.h}@ but was not initialized in any constructor. Subsequent code reads this uninitialized member, triggering the error.

@\textcolor{codepurple}{\textbf{\#\# Requirements}}@
A correct patch must satisfy the following requirements:
@\textcolor{codegreen}{\textbf{1.}}@  @\textbf{Initialize~the~Member:}@ @\code{redacted\_property\_name}@ must be initialized.
@\textcolor{codegreen}{\textbf{2.}}@  @\textbf{Correct~Default~Value:}@ Must be initialized to @\code{redactedValue}@ (which indicates an 'unset' state).
@\textcolor{codegreen}{\textbf{3.}}@  @\textbf{Robust~Implementation:}@ Use a C++11 in-class member initializer directly in the struct definition.
@\textcolor{codegreen}{\textbf{4.}}@  @\textbf{Correct~Location:}@ The change must be in @\code{some/redacted/path.h}@, not a workaround in the test.
\end{lstlisting}

\end{listing}

\begin{table}[t!]
  \centering
  \caption{When new raters use the same rubric, pairwise agreement improves to moderate/high agreement (Cohen's kappa).}
  \label{tab:same-rubric}
  \begin{tabular}{@{}llc@{}}
    \toprule
    \textbf{Rater} & \textbf{Rubric} & \textbf{Cohen’s kappa} \\
    \midrule
    Rater 1 and 2 & Rater 1’s rubric & 0.84 \\
    Rater 2 and 3 & Rater 2’s rubric & 0.67 \\
    Rater 3 and 1 & Rater 3’s rubric & 0.53 \\
    \bottomrule
  \end{tabular}
\end{table}

Following this template, the three raters developed golden rubrics and provided them to three new human raters who had no prior involvement in the study. These new raters were asked to assess the validity of the same 52 F2P patches. The new raters achieved substantial inter-rater agreement (Fleiss's kappa 0.66 as shown in Table~\ref{tab:same-rubric}), demonstrating that this template can help generate shared rubrics that effectively improve assessment consistency.

Overall, this preliminary study shows that using a templated rubric can effectively standardize the task and provide a reliable and consistent assessment of patch validity. These results motivate us to develop a framework to efficiently generate high-quality rubrics and apply rubric-guided assessment of patch validity at scale.

\section{Framework Overview}
\label{sec:design}

\begin{figure*}[t!]
    \centering
    \includegraphics[width=1.5\columnwidth]{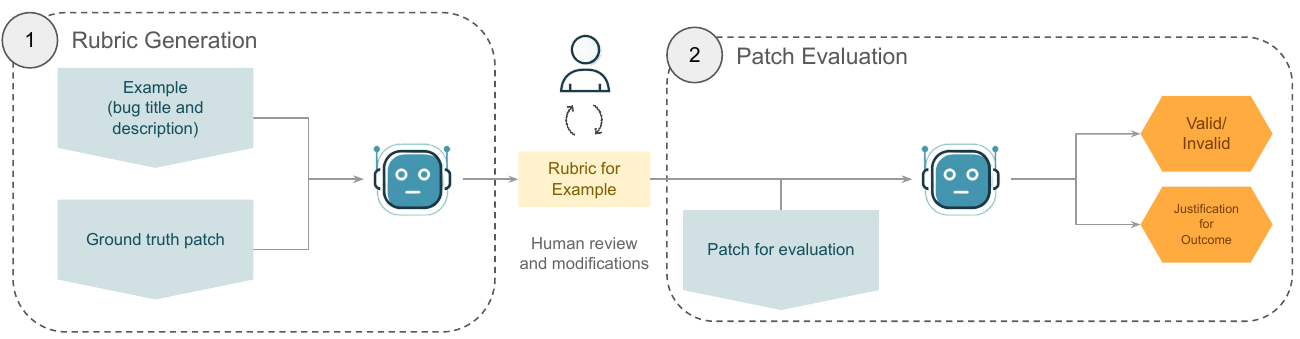}
    \caption{
    Overview of our two-stage framework. $\textcircled{1}$ \textbf{Rubric Generation}: An LLM first generates a per-bug rubric based on the bug's description and the ground truth patch. This rubric is reviewed and refined once by two human experts. $\textcircled{2}$ \textbf{Patch Evaluation}: The refined rubric is repeatedly used by an LLM judge to evaluate patches of the same bug. The LLM judge outputs a binary validity label and a natural language justification to the label, enabling both quantitative and qualitative analyses. 
    }
    \label{fig:llm-judge}
\end{figure*}

Based on the insights from our preliminary study, we develop a framework with the goal to provide a scalable and reliable offline evaluation for patch validity.
Figure \ref{fig:llm-judge} provides an overview of the two main stages of our framework:

\begin{enumerate}
    \item \textbf{Rubric Generation:} A one-time, human-in-the-loop process with LLM-based rubric generator to produce high-quality ``golden rubrics'' for patch evaluation.
    \item \textbf{Patch Evaluation:} A repeatable, automated process where an LLM judge applies the golden rubrics to assess patch validity.
\end{enumerate}

\subsection{Rubric Generation}
\label{sec:design:generation}
The rubric generation stage consists of two steps:

\paragraph{LLM-based draft rubric generation} We use an LLM to generate an example-specific rubric for each bug. Given the bug context (e.g., bug description) and its ground-truth fix as input, the LLM is instructed to generate the rubric draft that contains all the required sections listed in our predefined template (Listing \ref{lst:rubric-summary}), such as the root cause of the bug and a list of requirements for a valid fix.

\paragraph{Manual rubric refinement} Two human developers review and refine a draft rubric into a golden rubric.
One developer reviews and edits the draft rubric if needed, e.g., correcting inaccurate requirements, removing requirements overfitting to the ground-truth fix, and ensuring all elements required to fix the bug are captured. The developer also documents justifications for the edits. Another developer reviews and confirms all edits and the justifications. The final rubric will serve as the golden rubric. 

\subsection{Patch Evaluation}
\label{sec:design:judge}
Given the bug context (e.g., bug description), the patch in the unified diff format~\cite{unifieddiff}, and the golden rubric as input, the LLM judge is instructed to output its validity assessment.
The assessment consists of a ``thought'' section that captures its thinking steps, a binary label (``VALID'' or ``INVALID''), and a justification that captures its reasoning more concisely.
We store the LLM judge’s output label and justification for further analyses, such as qualitatively understanding why certain patches are accepted or rejected.

\subsection{Framework Formalization and Notations}
We now provide a simple formalization of our framework, with notation that will be used in the subsequent sections of this paper. We refer to the bug context (e.g., bug description) as $b$, the ground-truth fix (patch) as $f$, and a F2P patch as $p$. The two main stages of our framework can then be formalized into:
\begin{enumerate}
    \item An draft rubric $r$ is generated by the rubric generator $R$: $$r = R(b, f)$$
    Human developers, $H$, review and refine $r$ into the golden rubric $r_{\text{gold}}$: $$r_{\text{gold}} = H(r)$$
    \item The LLM judge $J_{\text{llm}}$ outputs a tuple $(v, t)$ on the patch $p$ of bug $b$ with golden rubric $r_{\text{gold}}$, where $v$ is the validity label and $t$ is justification for the label: $$(v, t) = J_{\text{llm}}(b, p, r_{\text{gold}})$$
\end{enumerate}

\section{Empirical Study Design}
\label{sec:evaluation}

We evaluate both the rubric generator and the LLM-as-a-Judge components of our framework to study its effectiveness. 
In this section, we describe the evaluation dataset, experimental setup and metrics used in our research questions.
We use Gemini 2.5 Pro as the LLM throughout our evaluation~\cite{comanici2025gemini}.

\subsection{Dataset} \label{dataset}
In this paper, we focus on sanitizer bugs, because (1) they can identify issues that can lead to security vulnerabilities, denial-of-service attacks, and other serious problems; and (2) they have structured bug reports that provide a consistent form of information (i.e., bug type, failing test, reproduction command) compared to human-reported bugs which may vary in detail.
These reasons make sanitizer bugs a promising area for high-impact and effective automated APR patch validation.
In our case, sanitizer bugs are automatically reported by a suite of sanitizer tools~\cite{googlesanitizers} that are run on a regular basis in Google's monorepo. 

We first collect 50 sanitizer bugs following prior bug curation process~\cite{rondon2025evaluating}.
We generate 20 patches for each bug using our APR agent~\cite{rondon2025evaluating}
and verify the F2P behavior of each patch by checking if the failing test from the bug report now passes after applying the patch.
We then randomly sample at most 3 unique F2P patches per bug; we ensure each sampled patch is unique by computing their hashes. Some bugs have fewer than 3 unique F2P patches generated, and 2 bugs have no F2P patches. Eventually, we obtained 115 unique F2P patches for 48 bugs.
Of these 48 bugs, we have 44 C++ bugs, 3 Go bugs, and 1 Java bug.
Table~\ref{tab:bug-types} summarizes the number of bugs and sampled patches by bug types.

\begin{table}[t!]
  \centering
  \caption{Bugs and sampled patches by bug types in our dataset.}
  \label{tab:bug-types}
  \begin{adjustbox}{width=0.8\columnwidth}
  \begin{tabular}{@{}llrr@{}}
    \toprule
    \textbf{Bug Type} & \textbf{\# Bugs} & \textbf{\# Patches} \\
    \midrule
    data\_race & 14 & 33 \\
    use\_of\_uninitialized\_value & 11 & 26 \\
    misaligned\_pointer\_use & 6 & 13 \\
    data\_race\_go & 3 & 8 \\
    fuzztest\_crash & 2 & 4 \\
    leak\_detected & 2 & 6 \\
    signed\_integer\_overflow & 2 & 5 \\
    stack\_use\_after\_scope & 2 & 6 \\
    use\_of\_uninitialized\_value & 2 & 5 \\
    hwasan\_tag\_mismatch & 1 & 2 \\
    invalid\_bool\_load & 1 & 2 \\
    invalid\_enum\_load & 1 & 3 \\
    null\_pointer\_use & 1 & 2 \\
    \midrule
    \textbf{Total} & 48 & 115 \\
    \bottomrule
  \end{tabular}
  \end{adjustbox}
\end{table}

\subsection{Research Questions}
\label{sec:evaluation:rq}

We investigate the following research questions in our study:
\begin{enumerate}
    \item \textbf{RQ1:} How effective is an LLM at generating reusable, high-quality rubrics for evaluating APR patches?
    \item \textbf{RQ2:} How well does an LLM-as-a-Judge align with human consensus on APR patch validity?
\end{enumerate}

\subsubsection{RQ1: Effectiveness of LLM in Rubric Generation}
\label{sec:evaluation:rq:generation}

We first use LLM-based rubric generator to produce draft rubrics for all 50 bugs.
A team of 6 authors of this work then perform rubric refinement (\S\ref{sec:design:generation}) on each draft rubric to produce a golden rubric.
To investigate this RQ, we study the changes between the 50 draft rubrics and their corresponding 50 golden rubrics with both quantitative metrics and qualitative analyses:

\begin{itemize}
    \item \textbf{Modification Rate:} The percentage of draft rubrics that were modified by human reviewers.
    \item \textbf{Normalized Edit Distance:} We compute the character-level Levenshtein distance between the draft rubric ($r$) and the golden rubric ($r_{\text{gold}}$), then normalize it by the draft rubric length:
    \[
    \text{NormalizedEditDistance} = \frac{\text{Levenshtein}(r, r_{\text{gold}})}{\text{length}(r)}
    \]
    This metric represents the proportion of the draft rubrics that was modified by the developer, and aims to quantify the magnitude of the relative change between the draft and the golden rubric.
    \item \textbf{Edit Type Analysis:} We analyze the nature of the changes by classifying edits as additions, deletions, or modifications.
    \item \textbf{Thematic Analysis:}
    We analyze the categories of edits by their semantic purposes (e.g., correcting inaccuracies, improving rubric generalization\Space{ by removing overfitting details}).
    To do so, we first provide the edits and justifications of these edits as input to LLM, and prompt the LLM to generate a initial set of categories.
    W\Space{e then conduct a manual review and final coding of the edit categories. During this process, w}e manually review and assign each distinct edit to one or more categories, while also iteratively refining the initial set of categories (e.g., via merging, splitting, or relabeling) to create a final, robust classification scheme that accurately reflects the edits' semantics.
\end{itemize}

\subsubsection{RQ2: Performance of LLM-as-a-Judge}

In this RQ, we aim to study the efficacy of LLM-as-a-Judge in assessing patch validity when it is guided by human-refined golden rubrics.
Specifically, we compare the patch validity labels produced by an LLM judge against the labels produced by human raters' consensus, and assess the agreements between an LLM judge and human consensus.

To establish a benchmark of patch validity from human consensus, three authors are assigned as raters to each of the 115 F2P patches. Working independently and without access to the LLM-as-a-Judge's outputs, each rater uses the golden rubrics to label each patch as ``VALID'' or ``INVALID''.
If the three raters produce different labels for a patch, they will discuss to reach consensus and establish a final, single label for the patch.

\paragraph{Human agreement} We measure the inter-rater reliability on the labels produced before the consensus discussions to check the agreement among human raters.

We use Krippendorff's alpha and a weighted average Cohen’s kappa (where the weight is based on the percentage of patches reviewed by each rater) to measure the inter-rater reliability among the human raters on their initial, independent ratings. This metric validates the consistency of the benchmark itself.

\paragraph{LLM-human agreement}
We use the LLM-as-a-Judge approach to produce validity labels for these patches, and compare its labels against those of the human consensus benchmark. 

We use accuracy and Cohen's kappa to measure the agreement between the LLM judge's labels and the human consensus. We also report precision, recall, and F1-Score of LLM judge for the VALID class by treating the human consensus as the ground truth.

\section{Empirical Study Results}
\label{sec:results}

In this section, we present results from studying the patch validation rubrics generated by our LLM-based rubric generator (\S\ref{sec:results:generation}), and results from studying the agreement between LLM-as-a-Judge and human consensus on patch validity (\S\ref{sec:results:judge}).

\subsection{RQ1: How Effective Is an LLM at Generating Patch Evaluation Rubrics?}
\label{sec:results:generation}

To study the quality of LLM-generated rubrics, we analyze the edits made by human developers during rubric refinement.

\subsubsection{Quantitative Analysis of Edits}
The LLM-based rubric generator produce draft rubrics with substantial details, with a median and mean length of 2,004 and 2,058 characters, respectively.

\begin{figure}[t!]
  \centering
  \includegraphics[width=0.8\columnwidth]{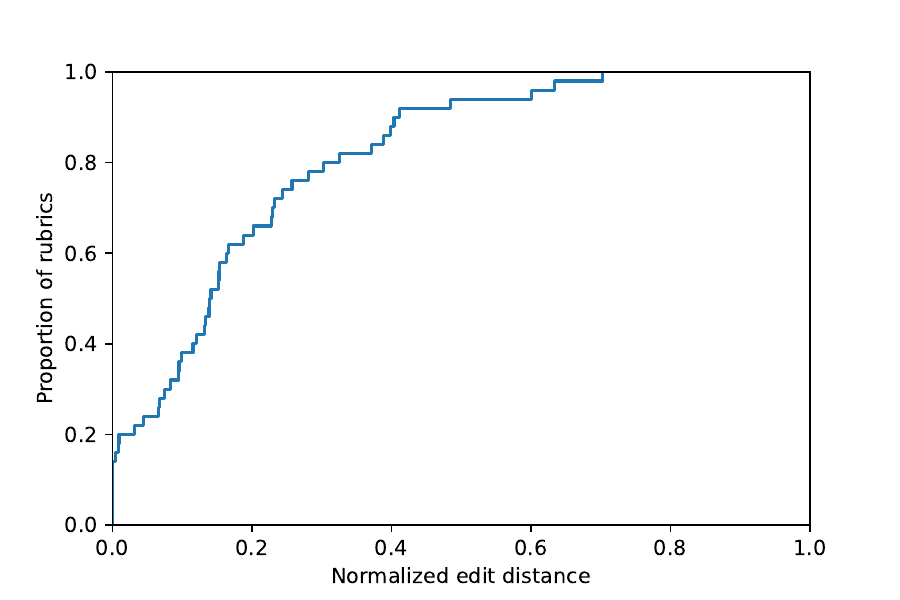}
  \caption{CDF of normalized edit distance on 50 rubrics. The median normalized edit distance is 14\% of the original content, while the maximum is 70\% of the original content.}
  \label{fig:edit-distance-dist}
\end{figure}

However, we found that 44 out of the 50 (88\%) draft rubrics needed manual revision before confirming them as golden rubrics. The magnitude of these revisions range from minor to considerable.
Specifically, the median absolute edit distance is 276 characters (average 385 characters). 
The median normalized edit distance was 0.14, indicating that 50\% of the draft rubrics required human revisions equivalent to 14\% of its original content, as shown in Figure~\ref{fig:edit-distance-dist}.
\Space{While some rubrics required only minor corrections, others demanded more substantial revisions, with t}The maximum observed normalized edit distance is 0.7.
Importantly, no LLM-generated draft rubric required a complete rewrite.

Overall, our results imply that LLM consistently provides a valuable scaffold, positioning the human developer's role for reviewing and refining rubrics rather than writing rubric from scratch.

\subsubsection{Edit Type Analysis}
We observe the diffs of the original and revised rubrics and classify the revisions into three types: deletion, addition, and modifications. 
We find that deletions were the most common\Space{ by a significant margin} (39 rubrics), followed by modifications (14 rubrics) and additions (3 rubrics).
The prevalence of deletions over additions suggests that LLM tends to generate overly verbose rubric drafts that include superfluous information, rather than drafts that are lacking details and incomplete.

\subsubsection{Edit Thematic Analysis}
We perform thematic analysis to understand underlying developer motivations for these edits.
Our analysis shows five primary categories of deficiencies of the LLM-generated drafts, as follows:
\begin{itemize}
    \item \textbf{Removing Superfluous Constraints (31 rubrics):} These edits eliminate overly general requirements not essential for a correct fix. Examples include ``changes must be minimal,''\Space{ adherence to specific coding styles not central to the fix\sam{probably not mentioning this because future work says to use llm judge for non-functional},} or mandating changes beyond the scope of the fix, such as adding unnecessary dependencies, tests, or comments.
    \item \textbf{Reducing Overfitting to Ground Truth (10 rubrics):} These edits generalize the rubrics to accept more correct fix implementations, rather than accepting only implementations that mirror the ground truth patch. These edits are critical for identifying more valid (but diverse) patches.
    \item \textbf{Correcting LLM Errors \& Hallucinations (9 rubrics):} These edits remove inaccurate diagnosis of the bug's root cause, hallucinated explanations, or other factually incorrect information in the LLM-generated rubrics.
    \item \textbf{Refining Scope \& Expectations (6 rubrics):} These edits refine fix scope, such as specifying which files should (or should not) be modified or ensuring that the core functional intent of a fix is preserved\Space{ in the evaluation criteria}.
    \item \textbf{Standardization (5 rubrics):} These edits improve consistency across all rubrics. Examples include standardizing rubric formatting, terminology (e.g., use ``primitive'' over ``fundamental''), and rubric structure. Standardizing rubrics could help LLM judge to have more predictable behavior.
\end{itemize}

While the LLM can generate promising draft rubrics as a strong scaffold, these deficiencies indicate that the human-in-the-loop process was essential to improve the LLM-generated draft rubrics before using them for patch evaluation.

\subsection{RQ2: How Well Does an LLM Judge Align with Human Consensus on Patch Validity?}
\label{sec:results:judge}

To compare LLM judge with human raters' judements, we establish a high-quality patch validity benchmark via human consensus, and then evaluate the performance of our LLM-as-a-Judge against it.

\subsubsection{Patch Validity Benchmark via Human Consensus}
Before any discussion to resolve disagreements, the raters achieved a perfect unanimous agreement on 81 out of 115 patches (70.4\%), and a Krippendorff’s alpha of 0.60 on all 115 patches. After the raters meet to resolve disagreements and reach consensus, the weighted average Cohen’s kappa between each human rater and the human consensus is 0.76, indicating moderate agreement.

Based on these results, we construct two benchmarks to evaluate LLM-as-a-Judge: a full benchmark $B_{\text{full}}$ containing all 115 patches\Space{ (44 VALID and 71 INVALID)}, and the clear benchmark $B_{\text{clear}}$ containing the 81 patches that have unanimous agreement prior to rater discussions\Space{ (35 VALID and 46 INVALID)}. 
The statistics of both benchmarks are further summarized in Table~\ref{tab:human-agreement}.

We also analyze the 34 patches where raters disagreed before reaching consensus and identify four disagreement themes:
\begin{itemize}
    \item \textbf{Overlooked Requirements (15 patches):} raters failed to identify or apply specific criteria detailed in the rubric.
    \item \textbf{Unnecessary Changes (9 patches):} raters disagreed on whether to accept code changes that go beyond the core issue the patch is supposed to address.
    \item \textbf{Rubric Ambiguity (8 patches):} unclear, incomplete, conflicting or differently interpreted rubric guidelines.
    \item \textbf{Correctness Assessment (2 patches):} raters disagreed on whether the patched code is functionally correct or introduces issues.
\end{itemize}

\begin{table}[t!]
  \centering
  \caption{Patch validity benchmarks.}
  \label{tab:human-agreement}
  \begin{adjustbox}{width=\columnwidth}
  \begin{tabular}{@{}lrr@{}}
    \toprule
    \textbf{Metric} & \textbf{$B_{\text{full}}$} & \textbf{$B_{\text{clear}}$} \\
    \midrule
    Number of bugs & 48 & 41 \\
    Number of patches & 115 & 81 \\
    Unanimous agreement & 70.4\% (81/115) & 100\% (81/81) \\
    Krippendorff’s alpha  & 0.60 & 1.0 \\
    Weighed average Cohen’s kappa & 0.76 & 1.0 \\
    Consensus: \# VALID patches & 44 & 35 \\
    Consensus: \# INVALID patches & 71 & 46 \\
    \bottomrule
  \end{tabular}
  \end{adjustbox}
\end{table}

\begin{table}[t!]
\centering
\caption{LLM judge effectiveness under different ablations. Using unrevised rubrics $r$, free-form rubrics $r_{\text{ff}}$, or ground truth (GT) patches all underperform our full approach $r_{\text{gold}}$.}
\label{tab:judge_comparison}
\begin{adjustbox}{width=\columnwidth}
\begin{tabular}{llcccc}
\toprule
& & \multicolumn{4}{c}{\textbf{Ablations}} \\
\cmidrule(lr){3-6}
\textbf{Metric} & \textbf{Benchmark} & \textbf{$r_{\text{gold}}$} & \textbf{$r$} & \textbf{$r_{\text{ff}}$} & \textbf{\shortstack{GT \\ patch}} \\ 
\midrule
\multirow{2}{*}{Cohen's kappa} & $B_{\text{full}}$ & \textbf{0.57} & 0.38 & 0.29 & 0.39 \\
 & $B_{\text{clear}}$ & \textbf{0.75} & 0.52 & 0.33 & 0.44 \\
\midrule
\multirow{2}{*}{Accuracy} & $B_{\text{full}}$ & \textbf{0.78} & 0.70 & 0.62 & 0.69 \\
 & $B_{\text{clear}}$ & \textbf{0.87} & 0.76 & 0.65 & 0.72 \\
\midrule
\multirow{2}{*}{Precision} & $B_{\text{full}}$ & \textbf{0.65} & 0.60 & 0.51 & 0.56 \\
 & $B_{\text{clear}}$ & \textbf{0.80} & 0.74 & 0.57 & 0.62 \\
\midrule
\multirow{2}{*}{Recall} & $B_{\text{full}}$ & \textbf{0.93} & 0.66 & 0.82 & 0.84 \\
 & $B_{\text{clear}}$ & \textbf{0.94} & 0.71 & 0.83 & 0.89 \\
\midrule
\multirow{2}{*}{F1-Score} & $B_{\text{full}}$ & \textbf{0.77} & 0.63 & 0.63 & 0.67 \\
 & $B_{\text{clear}}$ & \textbf{0.87} & 0.72 & 0.67 & 0.73 \\
\bottomrule
\end{tabular}
\end{adjustbox}
\end{table}

\subsubsection{LLM Judge Perfomance Against Human Consensus}
We evaluate the LLM-as-a-Judge ($J_{\text{llm}}$) using golden rubrics ($r_{\text{gold}}$) against both benchmarks. The bolded column in Table~\ref{tab:judge_comparison} shows the corresponding evaluation results\Space{, which demonstrate a moderate-to-substantial alignment with the consensus of human raters}.

The results indicate a moderate-to-substantial alignment on patch validity between the golden-rubric-guided LLM judge and the consensus of human raters.
Specifically, the primary alignment metric, Cohen's kappa, is 0.57 on $B_{\text{full}}$ and 0.75 on $B_{\text{clear}}$.
And the accuracy on $B_{\text{full}}$ and $B_{\text{clear}}$ are 0.78 and 0.87, respectively.

The LLM judge achieves a high recall of 0.93 on $B_{\text{full}}$ and 0.94 on $B_{\text{clear}}$, suggesting that it is highly effective at correctly identifying valid patches and minimizing false negatives.
We attribute this strong performance directly to the human refinement process on the rubrics.
Specifically, by generalizing the rubrics and removing overfitting to the ground truth patch (a prominent edit theme), the golden rubrics ($r_{\text{gold}}$) explicitly permit the judge to accept a wider range of correct, alternative solutions. The high recall of this framework also allows developers to use it to filter out INVALID patches, and reduce human effort in reviewing the remaining patches deemed VALID by the LLM judge.

Conversely, the precision of 0.65 on $B_{\text{full}}$ indicates that the judge is more prone to false positives, sometimes incorrectly labeling an invalid patch as valid.
However, a deeper analysis of these errors reveals a critical insight: the LLM judge's mistakes often overlap with cases that were also difficult for human experts. 
Of the 22 false positives generated by the LLM judge, 14 (63.6\%) occurred on patches that also generated disagreement among human raters (the 34 patches that required a consensus discussion).
This suggests that many of the LLM judge's mistakes occur on patches whose validity is controversial, which the human raters also find difficult to evaluate.
The higher precision (0.80) on the unanimously agreed patches ($B_{\text{clear}}$) further supports this observation.

We also categorize the mistakes LLM judge made on the 25 patches where it disagreed with the human consensus:
\begin{itemize}
\item     \textbf{Overlooked Requirements (10 patches):} The LLM judge failed to identify or apply specific criteria in the rubric.
\item \textbf{Unnecessary Changes (9 patches):} The LLM judge disagreed on whether to accept code changes that go beyond the core issue the patch is supposed to address.
    \item \textbf{Rubric Ambiguity (6 patches):} unclear, incomplete, conflicting or differently interpreted rubric guidelines.

\end{itemize}

Notably, LLM judge achieves a high negative predictive value (NPV) of 0.94 and 0.95 on $B_{\text{full}}$ and $B_{\text{clear}}$ respectively, indicating that its INVALID labels are highly reliable.
Developers can confidently refer to issues of these invalid patches identified by our framework for improving the evaluating APR system.

\section{Ablation Study}
\label{sec:ablation}

To better understand the impact of designs choices made in the rubric generation stage of our framework, we conduct several ablation studies and measure the resulting deviation from the human consensus benchmark.
We investigate the impact of (1) manual rubric refinement by comparing the LLM judge's performance with the golden rubric ($r_{\text{gold}}$) against the draft rubric ($r$); (2) the rubric template by comparing against a free-form rubric with no refinement ($r_\text{ff}$); and (3) the rubric itself by comparing against the common alternative of using the ground truth patch for patch evaluation. The performance of each configuration is summarized in Table~\ref{tab:judge_comparison}.

\subsection{Impact of Manual Rubric Refinement}
This study quantifies the value of the human-in-the-loop refinement step by comparing the performance of the LLM judge when using the final golden rubric ($r_{\text{gold}}$) versus the initial, draft rubric ($r$). 

Using the initial draft rubric ($r$) as a baseline, the LLM judge achieves only fair alignment (Cohen's kappa of 0.38 on $B_{\text{full}}$) with human raters. This is largely due to a poor recall of 0.66. This finding directly supports our analysis in RQ1: the initial, unedited rubrics are overfitted to the ground truth patch, and consequently misguide the LLM judge to reject a large number of valid alternative patches (i.e., produces many false negatives).

When human experts refine this initial draft into the golden rubric ($r_{\text{gold}}$), the performance improves dramatically. Recall improves from 0.66 to 0.93 on $B_{\text{full}}$, as the human-led generalizations (e.g., ``Reducing Overfitting to Ground Truth'' from RQ1) explicitly allow the judge to accept a wider range of correct solutions.
Precision is improved from 0.60 to 0.65 on $B_{\text{full}}$, as human edits correct the ``LLM Errors \& Hallucinations'' and ``Superfluous Constraints'' that misled the judge into accepting invalid patches.

Increase in precision and recall also boost Cohen's kappa from 0.38 to 0.57 on $B_{\text{full}}$. These results confirm that manual rubric refinement is indispensable for elevating this patch evaluation framework from a low-agreement baseline to a reliable instrument.

\subsection{Impact of Rubric Template}
Our framework's rubric generator, $R$, was designed to produce draft rubrics based on a manually-crafted, standardized template. To study the marginal contribution of this template, we compare performance of LLM judge on the unedited templated rubric ($r$) versus on the a ``template-free'' free-form rubric ($r_{\text{ff}}$), where mentions of template elements (root cause, requirements, examples of acceptable and unacceptable solutions) are removed from the rubric generation prompt.

Results in Table~\ref{tab:judge_comparison} show that the template is beneficial. Removing it causes Cohen's kappa to drop from 0.38 to 0.29 (i.e., fair agreement) on $B_{\text{full}}$.
This finding underscores the critical role of the rubric template: the structure the template imposes is not merely a formatting convention, but a key factor that guides the LLM-based rubric generator to reason about the root cause and encode different evaluation criteria in the rubric, which enables the downstream LLM judge to perform more reliable patch evaluation.

\subsection{Impact of Rubric versus Ground Truth Patch}

A common approach to use LLM-as-a-Judge for patch evaluation is to provide the ground truth patch as a reference, the task description, and the generated patch in the LLM prompt \cite{tong2024codejudge, zheng2023judging}. Table~\ref{tab:judge_comparison} (``GT Patch'') shows that this approach achieves Cohen’s kappa of 0.39 with human consensus on $B_{\text{full}}$, which is better than the free-form rubric, similar to draft rubric, and worse than our full approach.

\section{Discussion}
\label{sec:discussion}

\begin{figure}[t!]
  \centering
  \includegraphics[width=0.8\columnwidth]{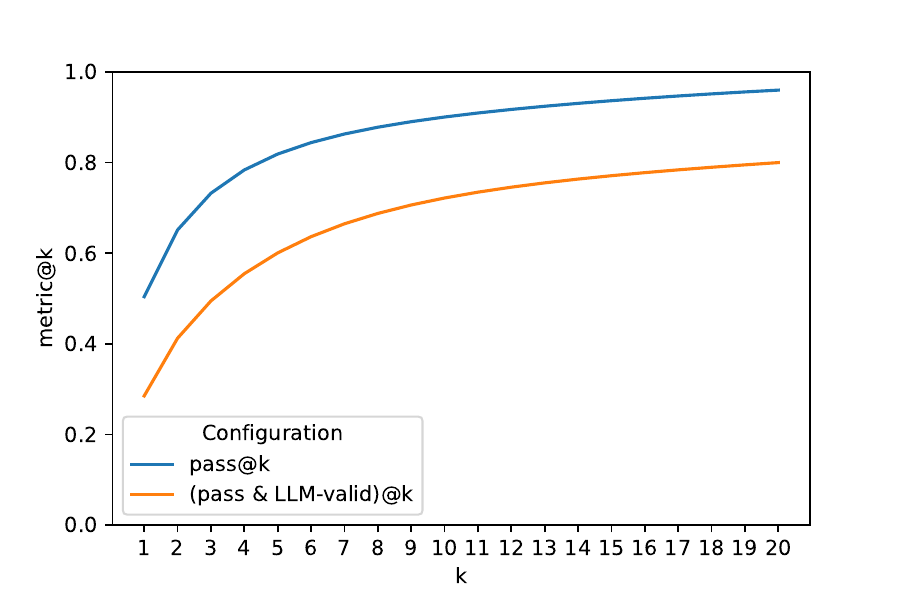}
  \caption{A 16-percentage-point drop from pass@20 to (pass \& LLM-valid)@20 due to LLM judge rejecting patches that violate requirements in rubric, fail to address root cause, are incomplete, or introduce new issues.}
  \label{fig:pass_valid_at_k}
\end{figure}

Figure~\ref{fig:pass_valid_at_k} compares pass@k and (pass \& LLM-valid)@k on the 1,000 patches (50 Sanitizer bugs $\times$ 20 generated patches) from \S\ref{dataset}. 
It shows a notable ``evaluation gap'' from Fail-to-Pass behavior to LLM-judged validity: at $k=20$, 96\% of the patches passed reproduction tests, while 80\% of the patches were judged valid. 

Out of 504 patches that passed bug reproduction tests, 219 (43.5\%) were deemed invalid by the LLM judge. We perform a thematic analysis on the judge's justifications to understand why they were rejected.
We cluster the justifications with an LLM, and identify these themes below:

\begin{itemize}

\item \textbf{Violating Specified Requirements in Rubric (60.6\%):}  Patches fail to adhere to specific guidelines, best practices, or architectural patterns outlined in the rubric. These violations often relate to coding style, design principles, memory management (e.g., requiring RAII/smart pointers vs. manual new/delete), solution scope (e.g., test-only changes vs. production code changes), adherence to idioms, performance expectations, or requirements for robustness and consistency. The patch might functionally fix the bug but does so in a manner that is explicitly disallowed in the rubric.

\item \textbf{Not Addressing Root Cause (26.6\%):} 
The patch either masks the symptom, provides a superficial workaround, misidentifies the problem, or applies a fix in an incorrect or irrelevant location. For example, sanitizer errors are suppressed instead of fixing undefined behavior, or production code is modified to work around a test data issue.

\item \textbf{Incomplete Implementation (7.3\%):} Patches do not fully address the scope of the problem; it leaves other parts of the same problem unresolved, fails to apply the fix to all affected instances or tests, or overlooks necessary related changes (e.g., missing dependencies, header includes).

\item \textbf{Introducing New Issues (5.5\%):} Patches that inadvertently introduce new problems, regressions, or vulnerabilities. These new issues can manifest as compilation errors, linking errors, new logical bugs, performance regressions, or the creation of an unstable state.
\end{itemize}

We also notice some of these patches were rejected by LLM judge due to the judge's misjudgement and inaccuracies from the rubric.
Specifically, judge justifications for 5 patches mention compilation errors, but these patches passed the reproduction tests. 
This misjudgement is because the rubric requires updating header and adding dependencies for the updated header. The rejected patch did not use the header and thus did not need to add such dependencies.

\section{Future Work}
\label{sec:future}

Building on our findings, we identify several promising directions for future research aimed at improving the efficiency and autonomy of our patch evaluation framework.

An immediate next step is to address the gap between the LLM judge and human consensus. We have identified the following opportunities for improving the system: (1) \textit{Rule of Minimalism}: implement a clear, universal rule that patches containing unnecessary changes should be classified as INVALID, which should resolve inconsistencies; and (2) \textit{Streamlined Rubric Review}: enable reviewers to assess rubrics against a set of diverse patches to confirm desired outcomes and reduce ambiguity.

We also plan to conduct a formal user study to quantify the efficiency gains from our framework. This study would measure the time and cognitive load required for an expert to refine an LLM-generated rubric compared to the effort of manually evaluating a large set of candidate patches from scratch.

Another significant avenue is to expand the judge output beyond binary correctness to non-functional aspects of code quality (e.g., maintainability, performance, security). The rubrics can also incorporate these criteria more considerably, leading to a multi-dimensional, holistic assessment of generated patches.

Our goal is to reduce reliance on human judgment and move toward a more autonomous APR evaluation ecosystem.
We envision a self-improvement cycle where the LLM judge automatically flags low-confidence assessments, solicits targeted human feedback only on these ambiguous cases, and then uses that feedback to automatically refine the rubrics. 
For example, we can guide the rubric generator with patch rejection justifications to produce higher-quality rubrics.

\section{Related Work}
\label{sec:related}

Test outcome has been a common metric for code correctness in program repair and translation evaluations \cite{pan2024lost, xia2023automated, joshi2023repair, chen2021evaluating, jimenez2023swe, austin2021program, xu2022systematic, ahmad2021avatar, roziere2020unsupervised}. As LLM-based APR systems are becoming more capable of generating high-quality patches \cite{comanici2025gemini, zhang2024autocoderover, bouzenia2024repairagent, yang2024swe}, developers need to further rely on manual assessment to determine patch validity (\S\ref{sec:introduction}). Studies of other code generation tasks, such as code summarization and text-to-code  \cite{lu2021codexglue, wang2021codet5, ahmad2021unified}, have used metrics derived from Natural Language Processing (NLP) literature to measure code structural and textual similarity \cite{papineni2002bleu, ren2020codebleu, eghbali2022crystalbleu, zhou2023codebertscore}, as well as code naturalness \cite{hindle2016naturalness}. However, these metrics do not capture the code validity that is critical to practical APR \cite{kulal2019spoc, roziere2020unsupervised, chen2021evaluating}.

More recently, researchers started employing LLM-as-a-Judge as a substitute for human experts to perform more tailored and scalable evaluations on increasingly-complex generative tasks \cite{zheng2023judging, tan2024judgebench}. In software engineering, recent studies explored the use of LLMs for evaluating and annotating coding task outputs such as generation \cite{tong2024codejudge, zhuge2024agent}, summarization \cite{crupi2025effectiveness, ahmed2025can}, and execution \cite{mundler2024swt, cheng2025agentic}. These studies have shown promising results with LLM judges, surpassing NLP metrics such as CodeBLEU~\cite{tong2024codejudge,ren2020codebleu} and aligning with human experts. Our work complements existing studies by further exploring the use of LLMs to not only judge, but also to generate bug-level fix requirements (captured by our rubric) for reliable and scalable APR evaluation in an industrial context.

\section{Threats to Validity}
\label{sec:threats}
\noindent \textbf{External Threats} concern the generalizability of our results to other contexts.
The study was limited to Google's internal monorepo and sanitizer bugs (C++, Java, Go). Findings may not apply to open-source projects or other bug classes, such as logical or UI errors.
We evaluated patches from only one internal APR system~\cite{rondon2025evaluating} and used only Geminin 2.5 Pro for LLM calls. Results may differ when applied to patches from other APR systems or using other LLMs.

\noindent \textbf{Internal Threats} concern potential confounding factors within our experimental setup.
Raters were experienced engineers but not the original code authors, potentially lacking deeper context. We mitigated this threat with consensus discussions for benchmark labeling.
The creation of golden rubrics involves human judgment, introducing potential variability. This was mitigated by having a second expert reviewer examine the final rubric.

\noindent \textbf{Construct Threats} concern whether our metrics accurately measure the concepts we claim to be studying.
Though our use of a binary VALID/INVALID classification is standard, we note that patches have varying degrees of acceptablility, and thus a binary judgment may, for example, overestimate the rate of patch rejection.

\section{Conclusion}
\label{sec:conclusion}

We introduce a human-in-the-loop framework centered on LLM-as-a-Judge to enable a more scalable and reliable offline evaluation for patches generated by APR systems.
Our approach is motivated by the insight that manual patch assessment suffers from low inter-rater reliability unless guided by a shared rubric.
Our framework operationalizes this insight by having human experts refine an LLM-generated rubric once per bug.
The rubrics can then be used repeatively by an LLM judge to evaluate patches for the same bug.

Our empirical study on 115 patches demonstrates the framework's potential as a reliable judge.
It achieves high recall (0.93) and proves highly reliable in its negative predictions (NPV of 0.94). When considering the 70.4\% of patches where human raters unanimously agreed, our judge reached substantial alignment (Cohen’s kappa 0.75) with high precision (0.8) and recall (0.94).
However, the judge's alignment on all the patches remains moderate (Cohen’s kappa 0.57).
The current results suggest that LLM-as-a-Judge can be reliable on bugs and patches where human annotators are reliable, and can be an effective automated screener that discards invalid, yet plausible patches that pass bug reproduction tests. %

\bibliographystyle{ACM-Reference-Format}
\bibliography{main.bbl}
\balance

\end{document}